\documentclass[narrowdisplay,onecolumn]{elsart3p}

 \usepackage{graphicx}

\usepackage[dvips]{color}

\usepackage{amssymb}

\makeatletter

\def\lsim{\compoundrel<\over\sim}
\def\compoundrel#1\over#2{\mathpalette\compoundreL{{#1}\over{#2}}}
\def\compoundreL#1#2{\compoundREL#1#2}
\def\compoundREL#1#2\over#3{\mathrel
	{\vcenter{\hbox{$\m@th\buildrel{#1#2}\over{#1#3}$}}}}
\makeatother
\newcommand{\bfi}[1]{\mbox{\boldmath $#1$}}

\begin{document}

\begin{frontmatter}
\hfill {\small UTCCS-P-43, TKYNT-08-07, UTHEP-564}

\title{ Hyperon-Nucleon Force from Lattice QCD }

\author[1]{Hidekatsu Nemura\corauthref{cor}\thanksref{presentaddr}},
 \corauth[cor]{Corresponding author.}
 \thanks[presentaddr]{Present address:
 Strangeness Nuclear Physics Laboratory,
 Nishina Center for Accelerator-Based Science, RIKEN,%
 Wako  %
 351-0198, Japan 
 }
 \ead{nemura@riken.jp}
\author[2]{Noriyoshi Ishii},
\author[3,4]{Sinya Aoki}, 
and 
\author[5]{Tetsuo Hatsuda}

\address[1]{Advanced Meson Science Laboratory,
        Nishina Center for Accelerator-Based Science, RIKEN,%
	Wako  %
        351-0198, Japan 
 }
\address[2]{Center for Computational Sciences,
	University of Tsukuba, %
	Tsukuba 305-8571, Japan 
 }
\address[3]{Graduate School of Pure and Applied Sciences,
	University of Tsukuba, %
	Tsukuba 305-8571, Japan 
 }
\address[4]{Riken BNL Research Center, Brookhaven National Laboratory, %
	Upton, New York 11973, USA 
 }
\address[5]{Department of Physics, The University of Tokyo, %
	Tokyo 113-0033, Japan
 }

\begin{abstract}
 We calculate  
 potentials between a proton and a $\Xi^0$
 (hyperon with  strangeness $-2$) 
 through the equal-time Bethe-Salpeter wave function,
 employing 
 quenched lattice QCD simulations with
 the plaquette gauge action and the
  Wilson quark action
 on  (4.5 fm)$^4$ lattice at the lattice spacing $a \simeq 0.14$ fm.
 The $ud$ quark mass in our study 
 corresponds to $m_{\pi}\simeq 0.37$ and 0.51 GeV, while
 the $s$ quark mass corresponds to the physical value of $m_K$.
 The central $p\Xi^0$ potential has a 
 strong (weak) repulsive core in the $^1S_0$ 
 ($^3S_1$) channel for $r \lsim 0.6$ fm,  while 
 the potential has attractive well 
 at medium and long distances  ($0.6$ fm $\lsim r \lsim 1.2$ fm) 
 in both channels.
 The sign of the $p\Xi^0$ scattering length and
 its quark mass dependence
 indicate a net attraction in both channels at low energies.
\end{abstract}

\end{frontmatter}

\section{Introduction}

Modern nucleon-nucleon ($NN$) potentials \cite{Machleidt:2001rw}
give  successful description of the $NN$ scattering data
at low energies
and have been used for precision calculations in  
light nuclei \cite{Pieper:2007ax}.  
 Similarly,  hyperon-nucleon ($YN$) and hyperon-hyperon ($YY$) 
potentials
  in the coordinate space
 are known to be quite useful and important for studying  the 
   properties of  hypernuclei \cite{Hashimoto:2006aw}
   as well as the equation of state for 
   hyperonic matter in  neutron stars \cite{Ishizuka:2008gr}. 
 However, there are still  large experimental uncertainties
 in $YN$ and $YY$ interactions at present, because 
 the short life-time of the hyperons makes
 the scattering experiments difficult.
 Accordingly,  phenomenological  $YN$ and $YY$ potentials
 are not well constrained from data even under some theoretical guides
 \cite{ESC04,Oka:2000wj,Fujiwara:2006yh,Arisaka:2005jb,Kondo,Haidenbauer:2007ra}.

In such a situation, it would be desirable to analyse
 hyperon interactions on the basis of the first principle lattice QCD simulations.
 Studies along this line for nucleon interactions was 
  initiated in Ref.\cite{Fukugita:1994ve}, where the $NN$ scattering lengths were
  extracted from the quenched simulations by using the 
    L\"{u}scher's finite volume method \cite{Luscher}. 
   The same method was applied later to the
  (2+1)-flavor simulations with the mixed action \cite{NPLQCD}.
  The scattering length of  $\Lambda N$ system was first examined
  in \cite{Muroya:2004fz} in quenched simulations  on a small lattice box.
  Subsequently, $\Lambda N$ and $\Sigma N$ scattering lengths were studied
  in (2+1)-flavor simulations with the mixed action \cite{Beane}.
 Such low-energy scattering parameters calculated in lattice QCD would be 
 valuable inputs to construct phenomenological $YN$ and $YY$ potentials to be used
 for studying hypernuclei and hyperonic matter. 
 
   Recently, an alternative but closely related approach to \cite{Luscher}
    has been proposed to define the $NN$ potential from  lattice QCD 
  \cite{Ishii:2006ec,Ishii:2007xz,Aoki:2008hh}. Since 
  the potential is not a direct physical observable,
  one cannot make quantitative comparison of this 
  lattice $NN$ potential with the phenomenological $NN$ potentials.
    However,  they are both designed to reproduce the correct scattering phase shifts,
  and  their spatial structures are found to have common features
   \cite{Ishii:2006ec}; 
 the attraction at long and intermediate distances 
  and  the strong repulsion at short distance in the $S$-wave channel \cite{jastrow}.

   The purpose of the present paper
  is to report our first attempt to apply  the above approach to 
 the hyperon-nucleon systems. 
  For the $NN$ potential, 
 only two channels (isovector and isoscalar)
 exist in the flavor SU(2) space,  i.e., 
${\bf 2}\otimes{\bf 2}={\bf 3}\oplus{\bf 1}$.
Including the strange quark extends this to
$
{\bf 8}\otimes{\bf 8}=
 {\bf 27}\oplus {\bf 10^*}\oplus{\bf 1}\oplus{\bf 8}\oplus
 {\bf 10}\oplus{\bf 8}
$ in the flavor SU(3) space.
 The isovector (isoscalar) channel in the $NN$ sector is 
assigned to be a subset in the ${\bf 27}$-plet 
(${\bf 10^*}$-plet) representation. 
The potentials in newly arising channels 
are hardly determined from experiments so far.

In this paper, we focus on the $N\Xi$ potential 
in the isovector ($I=1$) channel as a first step.
(A preliminary account of this system has been reported in \cite{Nemura:2007cj}.)
There are two main reasons for picking up this channel:
(i) Theoretically, it is the simplest generalization of the $NN$ system;
 $p\Xi^0$ is obtained from $pn$ by replacing 
 the $d$-quarks in the neutron by the $s$-quarks. Also
 it is a channel  which do not have strong decay into
  other $YN$ systems.\footnote{Note that
 $N\Xi$ in the isoscalar ($I=0$) channel
 is above the $\Lambda\Lambda$ threshold.} 
(ii)  Experimentally, not much information has been available 
 on the $N\Xi$ interaction except for a few studies;
 a recent report gives the upper limit of elastic and
 inelastic cross sections \cite{XN_Ahn},
 and earlier publications suggest 
  weakly attractive $\Xi$-nucleus potential \cite{Nakazawa}.
The $\Xi$-nucleus interaction will be soon 
 studied  as one of the day-one experiments 
 at J-PARC \cite{JPARC} via $(K^-,K^+)$ reaction
  with nuclear target.  

This paper is organized as follows. 
In Section~\ref{sec:formulation}, we describe the basic formulation 
to derive the $p\Xi^0$ potential through the Bethe-Salpeter amplitude 
measured in the lattice QCD simulations. 
Our lattice setup is explained in Section~\ref{sec:setup}. 
In Section~\ref{sec:results}, we show numerical results
 of the potentials in the spin-singlet and spin-triplet channels
  and their quark mass dependence.
 We also show the estimate of the scattering lengths in these channels.
 Section~\ref{sec:summary} is devoted to summary and concluding remarks.

\section{Basic formulation}
\label{sec:formulation}

Our methodology to obtain the baryon potentials is
along the lines of
 \cite{Ishii:2006ec,Ishii:2007xz,Aoki:2008hh}.  (See also  \cite{Luscher,CP-PACS} for
  the seminal attempts to introduce similar notion of the potential.)  
We consider the low-energy $N$-$\Xi$ scattering 
and start with an effective Schr\"{o}dinger equation
 for the equal-time Bethe-Salpeter (BS)
 wave function $\phi(\vec{r})$ obtained from the 
Lippmann-Schwinger equation \cite{Aoki:2008hh,AHI_full};  
\begin{equation}
 -{1\over 2\mu}\nabla^2 \phi(\vec{r}) +
  \int U(\vec{r},\vec{r}^\prime)
  \phi(\vec{r}^\prime) d^3r^\prime  =
  E \phi(\vec{r}).
  \label{effSchrEq}
\end{equation}
Here $\mu=m_{N}m_{\Xi}/(m_{N}+m_{\Xi})$ and 
$E\equiv k^2/(2\mu)$ are the reduced mass of the $N\Xi$ system and 
the non-relativistic energy in the center-of-mass frame, respectively. 
Eq.~(\ref{effSchrEq}) is derived from QCD 
by adopting local three-quark operator as the nucleon interpolating 
operator.  It can be shown that the non-local potential $U(\vec{r},\vec{r}^\prime)$
is  energy independent. Also, it is unique given the interpolating operator 
  as long as  the BS-amplitude for all energies is known  \cite{Aoki:2008hh}.
 If we take other  nucleon interpolating operator,
 both the BS wave function and the non-local potential change
 their spatial structure without affecting the 
 observables such as the binding energy and the phase shift.   
 At low energies, the nonlocality of the  potential can be expanded as 
   \cite{TW67}
\begin{equation}
U(\vec{r},\vec{r}^\prime)=
 V_{N\Xi}(\vec{r},\vec{\nabla})\delta(\vec{r}-\vec{r}^\prime),
\end{equation}
where the dimensionless expansion parameter is
  $\nabla/M$ with  $M$ being the typical scale of the strong interaction
  such as the excitation energy of a single baryon.
The general expression of the potential $V_{N\Xi}$ 
is known to be \cite{JJdeSwart1971} 
\begin{eqnarray}
 V_{N\Xi} &=&
  V_0(r)
  +V_\sigma(r)(\vec{\sigma}_N\cdot\vec{\sigma}_\Xi)
  +V_\tau(r)(\vec{\tau}_N\cdot\vec{\tau}_\Xi)
  +V_{\sigma\tau}(r)
   (\vec{\sigma}_N\cdot\vec{\sigma}_\Xi)
   (\vec{\tau}_N\cdot\vec{\tau}_\Xi)
   \nonumber \\
 &&
  +V_T(r)S_{12}
  +V_{T\tau}(r)S_{12}(\vec{\tau}_N\cdot\vec{\tau}_\Xi)
  +V_{LS}(r)(\vec{L}\cdot\vec{S}_+)
  +V_{LS\tau}(r)(\vec{L}\cdot\vec{S}_+)(\vec{\tau}_N\cdot\vec{\tau}_\Xi)
  \nonumber \\
 &&
  +V_{ALS}(r)(\vec{L}\cdot\vec{S}_-)
  +V_{ALS\tau}(r)(\vec{L}\cdot\vec{S}_-)(\vec{\tau}_N\cdot\vec{\tau}_\Xi)
  +{O}(\nabla^2).
  \label{GenePot}
\end{eqnarray}
Here
$S_{12}=3(\vec{\sigma}_1\cdot\vec{n})(\vec{\sigma}_2\cdot\vec{n})-\vec{\sigma}_1\cdot\vec{\sigma}_2$
is the tensor operator with $\vec{n}=\vec{r}/|\vec{r}|$,
$\vec{S}_{\pm}=(\vec{\sigma}_1 \pm \vec{\sigma}_2)/2$  are %
symmetric ($+$) and antisymmetric ($-$) spin operators,
$\vec{L}=-i\vec{r}\times\vec{\nabla}$ is the orbital %
angular momentum operator, 
and 
$\vec{\tau}_N$ ($\vec{\tau}_{\Xi}$) is isospin operator 
for $N=(p,n)^{\rm T}$ ($\Xi= (\Xi^0,\Xi^-)^{\rm T}$). 
We note that the antisymmetric spin-orbit forces 
($V_{ALS}$ and $V_{ALS\tau}$) 
do not arise %
in the $NN$ case because of the identical nature of the 
nucleon within the isospin symmetry.

According to the above expansion, 
the wave function should be classified by the total isospin $I$, 
the total angular momentum 
$\vec{J}=\vec{L}+\vec{S}_+$ and the parity. 
A particular spin (isospin) projection is %
made in terms of 
$\vec{\sigma}_N\cdot\vec{\sigma}_\Xi$
($\vec{\tau}_N\cdot\vec{\tau}_\Xi$), e.g., 
for the isospin projection we have 
$P^{(I=0)}=(1-\vec{\tau}_N\cdot\vec{\tau}_\Xi)/4$ 
and 
$P^{(I=1)}=(3+\vec{\tau}_N\cdot\vec{\tau}_\Xi)/4$. 
  
The equal-time BS wave function for $(I,I_z)=(1,1)$ and $L=0$ ($S$-wave) %
 on the lattice is obtained  by
\begin{eqnarray}
 \phi(\vec{r}) &=&
  {1\over 24} \sum_{{ R}\in O} {1\over L^3} \sum_{\vec{x}}
  P^\sigma_{\alpha\beta} 
  \left\langle 0
   \left|
    p_\alpha({ R}[\vec{r}]+\vec{x})
    \Xi^0_\beta(\vec{x})
   \right| p \Xi^0 ; k
  \right\rangle,
  \\
  p_\alpha(x) &=&
  \varepsilon_{abc} \left(
		     u_a(x) C \gamma_5 d_b(x)
		    \right)
  u_{c\alpha}(x),
  \label{proton}
  \\
  \Xi^0_\beta(y)  &=&
  \varepsilon_{abc} \left(
		     u_a(y) C \gamma_5 s_b(y)
		    \right)
  s_{c\beta}(y),
  \label{Xi}
\end{eqnarray}
where $\alpha$ and $\beta$ denote the Dirac indices, 
$a$, $b$ and $c$ the color indices, and 
$C=\gamma_4\gamma_2$ the charge conjugation matrix.
The summation over ${ R}\in O$ is taken for cubic transformation 
group to project out the $S$-wave state.\footnote{
 Due to the periodic boundary condition, this projection cannot remove 
 the higher orbital components with $L \ge 4$, which however are expected to be small in the ground state. }
The summation over $\vec{x}$ is to select the state with  zero total momentum. 
Here we take local field operator $p_\alpha(x)$ and $\Xi^0_\beta(y)$ 
for the proton and $\Xi^0$. 
  The wave function and the 
  potential (or equivalently the off-shell
   behavior of the scattering amplitude) 
   depend on the choice of  interpolating operators.
  This  is the situation  common to any field theories.
  In this paper, we focus exclusively on the local operators
   as introduced 
  above and leave further discussions on the operator
   dependence to \cite{AHI_full}.
We take the upper components of the Dirac indices 
to construct the spin singlet (triplet) channel by 
$P^{(\sigma=0)}_{\alpha\beta}=(\sigma_2)_{\alpha\beta}$
($P^{(\sigma=1)}_{\alpha\beta}=(\sigma_1)_{\alpha\beta}$). 
 The BS wave function $\phi(\vec{r})$ 
is understood as a probability amplitude to find 
``nucleon-like'' three-quarks located at point $\vec{x}+\vec{r}$ and
``$\Xi$-like'' three-quarks located at point $\vec{x}$.
  Our BS wave function
 has information not only of the
elastic amplitude $N\Xi\rightarrow N\Xi$ but also of the
inelastic amplitudes:  
The inelastic effects are localized
in coordinate space at low energies and do not affect
 the asymptotic behavior of $\phi(\vec{r})$ and hence the phase shift.

In the actual simulations, the BS wave function is obtained from the 
four-point correlator,
\begin{eqnarray}
\label{eq:4-point}
 F_{p\Xi^0}(\vec{x}, \vec{y}, t; t_0) &=&
  \left\langle 0
   \left|
    p_\alpha(\vec{x},t)
    \Xi^0_\beta(\vec{y},t)
    \overline{ J}_{p \Xi^0}(t_0)
   \right| 0 
  \right\rangle
  \\
 &=&
  \sum_n A_n
  \left\langle 0
   \left|
    p_\alpha(\vec{x})
    \Xi^0_\beta(\vec{y})
    \right| E_n
  \right\rangle
  {\rm e}^{-E_n(t-t_0)}.
\label{eq:BSamp}
\end{eqnarray}
Here $\overline{ J}_{p\Xi^0}(t_0)$ is a wall source located at
$t=t_0$,
which is defined by 
${J}_{p\Xi^0}(t_0)=
P_{\alpha\beta}^{\sigma} 
{p}_{\alpha}(t_0)
{\Xi}_{\beta}^{0}(t_0)$ with 
${p}_{\alpha}(t_0)=\sum_{\vec{x}_1,\vec{x}_2,\vec{x}_3}
\varepsilon_{abc}(u_a(\vec{x}_1,t_0)C \gamma_5
d_b(\vec{x}_2,t_0))u_{c\alpha}(\vec{x}_3,t_0)$ and 
${\Xi}_{\beta}^{0}(t_0)=\sum_{\vec{y}_1,\vec{y}_2,\vec{y}_3}
\varepsilon_{abc}(u_a(\vec{y}_1,t_0)C \gamma_5
s_b(\vec{y}_2,t_0))s_{c\beta}(\vec{y}_3,t_0)$. 
 The eigen-energy and the eigen-state of the
  six quark system are denoted by $E_n$ and $| E_n \rangle$, respectively,
  with the matrix element $A_n(t_0)=\langle E_n |\overline{ J}_{p\Xi^0}(t_0)|0\rangle$. 
For $t-t_0 \gg 1$, the $F_{p\Xi^0}$ %
and hence the wave function $\phi$ are dominated 
by the lowest energy state. 

The lowest energy state $| E_0\rangle$ 
 created by the wall source $\overline{ J}_{p \Xi^0}(t_0)$
 contains not only the $S$-wave $N\Xi$ component but also 
 the components which can mix by QCD dynamics, such as the
 $D$-wave due to the tensor force and $\Lambda\Sigma$ state 
 due to the rearrangement of the quarks inside baryons.
 In principle, these components can be disentangled
 by preparing appropriate operator sets for the sink.
 Study along this line to extract the mixing between the
  $S$-wave and the $D$-wave in low energy $NN$ interaction 
  was put forward recently in \cite{IshiiCHIRAL07Proc}. 
   In the present paper, instead of
  making such decomposition, we 
 define  an  effective central potential $V_{\rm C}(r)$
 for the $S$-wave component according to our previous works
 \cite{Ishii:2006ec,Ishii:2007xz,Aoki:2008hh}: 
 \begin{equation}
 V_{\rm C}(r) = E +
  {1\over 2\mu}{\vec{\nabla}^2{\phi}(r)\over {\phi}(r)}.
  \label{EffCPot}
\end{equation}
Such a potential in the $I=1$ sector 
has only the spin dependence as
$V_{\rm C}^{(I=1)}(r) = \tilde{V}_0(r) + \tilde{V}_{\sigma}(r)
 \left(  \vec{\sigma}_N \cdot \vec{\sigma}_{\Xi} \right)$ where 
  effects of the tensor force and coupled-channel effects 
 are implicitly taken into account in $\tilde{V}_{0,\sigma}$. 
 
It is in order here to make some remarks on 
 the  potential we have defined.  
 \begin{enumerate}
  \item[(i)] 
The description by the energy-independent non-local potential
   $U(\vec{r},\vec{r}^\prime)$  is
   equivalent to that by the momentum-dependent local  
    potential, $V(\vec{r},\vec{\nabla})$. 
     The potential
   $V_{\rm C}(r)$  in Eq.(\ref{EffCPot}) is the
    leading order term of the derivative expansion of 
    $V(\vec{r},\vec{\nabla})$ and can be determined
     by the BS wave function at $E \simeq 0$.
    Higher order terms can be extracted successively by 
   the BS wave functions with different $E$. 
   It was recently reported that momentum-dependence of   
   $V(\vec{r},\vec{\nabla})$ is small at least 
   up to the center of mass momentum $p=250$ MeV for the $NN$ 
    system
   \cite{Aoki:0812_0673}.
  \item[(ii)]   
	     The non-local potential $U(\vec{r},\vec{r}^\prime)$
 has one-to-one correspondence to the baryon interpolating 
 operator adopted in defining the Bethe-Salpeter amplitude.
 Different choices of the interpolating operator would
 give different BS wave function and the baryon-baryon potential, although 
 the phase shifts and binding energies are unchanged.
An advantage of working directly in lattice QCD is that
  we can unambiguously trace the relation between the 
  baryon interpolating operator and the BS wave function or equivalently
   the associated potential.
\end{enumerate}

\section{Lattice setup and hadron masses}
\label{sec:setup}

We use the plaquette gauge action and the
Wilson fermion action with the gauge coupling
$\beta=5.7$ on the $32^3\times 32$ lattice.
The heatbath algorithm combined with the overrelaxation is used 
to generate quenched gauge configurations. 
After skipping 3000 sweeps for thermalization,
measurement is made for every 200 sweep. 
The Dirichlet (periodic) boundary condition is imposed for
quarks in the temporal (spatial) direction. 
The wall source is placed at $t_0=5$ 
with the Coulomb gauge fixing. 
These setup are same as our previous  calculation  
for the $NN$ potential \cite{Ishii:2006ec}.

Masses of pseudo-scalar and vector mesons obtained at hopping parameter
$(\kappa_1, \kappa_2)$ are fitted as
\begin{equation}
 (m_{\rm ps}a)^2 = {B\over 2} \left(
			   {1\over \kappa_1} - {1\over \kappa_c} 
			  \right)
 + {B\over 2}\left(
	      {1\over \kappa_2} - {1\over \kappa_c} 
	     \right),
 \qquad
 m_{\rm v}a = C + {D\over 2} \left(
				{1\over \kappa_1} - {1\over \kappa_c} 
			       \right)
 + {D\over 2}\left(
	      {1\over \kappa_2} - {1\over \kappa_c} 
	     \right).
\end{equation}
for $\kappa_i = 0.1678, 0.1665, 0.1640$ ($i=1,2$). 
The fit gives $\kappa_{\rm c}=0.16930(1)$
for the critical hopping parameter.
From $m_\pi = 135$ MeV and $m_\rho = 770$ MeV,
we determine the hopping parameter for physical $ud$ quarks and the lattice spacing
as  $\kappa_{\rm phys}=0.16910(1)$ and  $a=0.1416(9)$~fm ($1/a = 1.393(9)$ GeV), respectively, while
the hopping parameter for the strange quark mass is given by
 $\kappa_s=0.16432(6)$ from $m_K=494$ MeV. 
 A corresponding lattice volume is $(4.5 {\rm fm})^4$, which is 
  large enough to accommodate two baryons.

In order to check the thresholds of  two baryon systems 
with strangeness $S=-2$ 
(${\Lambda\Lambda}$,${N\Xi}$,${\Lambda\Sigma}$ and ${\Sigma\Sigma}$),
 we calculate the two-point correlator, 
$C(t;t_0)=\sum_{\vec{x}}\langle 0|{ B}_\alpha(\vec{x},t)
\overline{ J}_{B_\alpha}(t_0)|0\rangle$, 
for the octet baryons (${ B}=N,\Xi,\Lambda,\Sigma$), 
and 
$J_{B_\alpha}(t_0)$ is the wall-source for $B$. 
The interpolating fields for $\Lambda$ and $\Sigma^+$
employed in this work are 
\begin{equation}
 \Lambda_\alpha(x)=
  \varepsilon_{abc} \left\{
		     \left(
		      d_a(x) C \gamma_5 s_b(x)
		     \right) u_{c\alpha}(x)
		     + \left(
			s_a(x) C \gamma_5 u_b(x)
		       \right) d_{c\alpha}(x)
		     - 2 \left(
			  u_a(x) C \gamma_5 d_b(x)
			 \right) s_{c\alpha}(x)
		    \right\},
\end{equation}
\begin{equation}
 \Sigma^+_\beta(y)=
  - \varepsilon_{abc} \left(
		     u_a(y) C \gamma_5 s_b(y)
		    \right)
  u_{c\beta}(y).
\end{equation}
Table~\ref{masses} lists the meson and baryon masses measured 
at two values of $\kappa_{ud}$, 0.1665 and 0.1678, 
with fixed $\kappa_s=0.1643$.
At $\kappa_{ud}=0.1678$ corresponding to $m_{\pi} \simeq 368$ MeV,
17 exceptional configurations are excluded from totally 1300 gauge configurations
for the average, while 
no such exceptional configuration appears at  $\kappa_{ud}=0.1665$ 
($m_{\pi} \simeq 511$ MeV).
 As seen in Table~\ref{masses}, the present results for the baryon masses 
 are consistent with  the experimentally observed ordering of the two-baryon thresholds
 in the strangeness $-2$ sector: 
 $E_{\rm th}(\Lambda\Lambda) < E_{\rm th}(N\Xi) < E_{\rm th}(\Lambda\Sigma)<
 E_{\rm th}(\Sigma\Sigma)$. 
This warrants that $N\Xi$ in the $I=1$ channel treated in this paper
is indeed the lowest energy scattering state.

\begin{table}[b]
 \centering \leavevmode 
 \begin{tabular}{cccccccccccc}
  \hline \hline
  $\kappa_{ud}$ & $\kappa_s$ & $N_{\rm conf}$ & 
  $m_\pi$ & $m_\rho$ & $m_K$ & $m_{K^\ast}$ & $m_{\phi}$ &
  $m_p$ & $m_{\Xi^0}$ & $m_\Lambda$ & $m_{\Sigma^+}$ \\
  \hline
  0.1665 & 0.1643 & 1000 &
  511.2(6) & 861(2) & 605.3(5) & 904(2) & 946(1) &
  1300(4) & 1419(4) & 1354(4) & 1375(4) \\
  0.1678 & 0.1643 & 1283 & 
  368(1) & 813(4) & 554.0(5) & 884(2) & 946(1) &
  1167(7) & 1383(6) & 1266(6) & 1315(6) \\
  \hline
       &        &  Exp.    & 
   135   &  770   & 494  & 892   & 1019 &
  940    & 1320   & 1116 & 1190        \\
 \hline \hline 
  \end{tabular}
 \caption{Hadron masses in the unit of MeV calculated 
 for the hopping parameters $\kappa_{ud}$ and $\kappa_s$, 
 with the number of gauge configurations $N_{\rm conf}$.}
 \label{masses}
\end{table}

\section{Results of $N\Xi$ interaction}
\label{sec:results}

\subsection{BS wave function}

\begin{figure}[t]
 \centering \leavevmode
 \includegraphics[width=.480\textwidth]{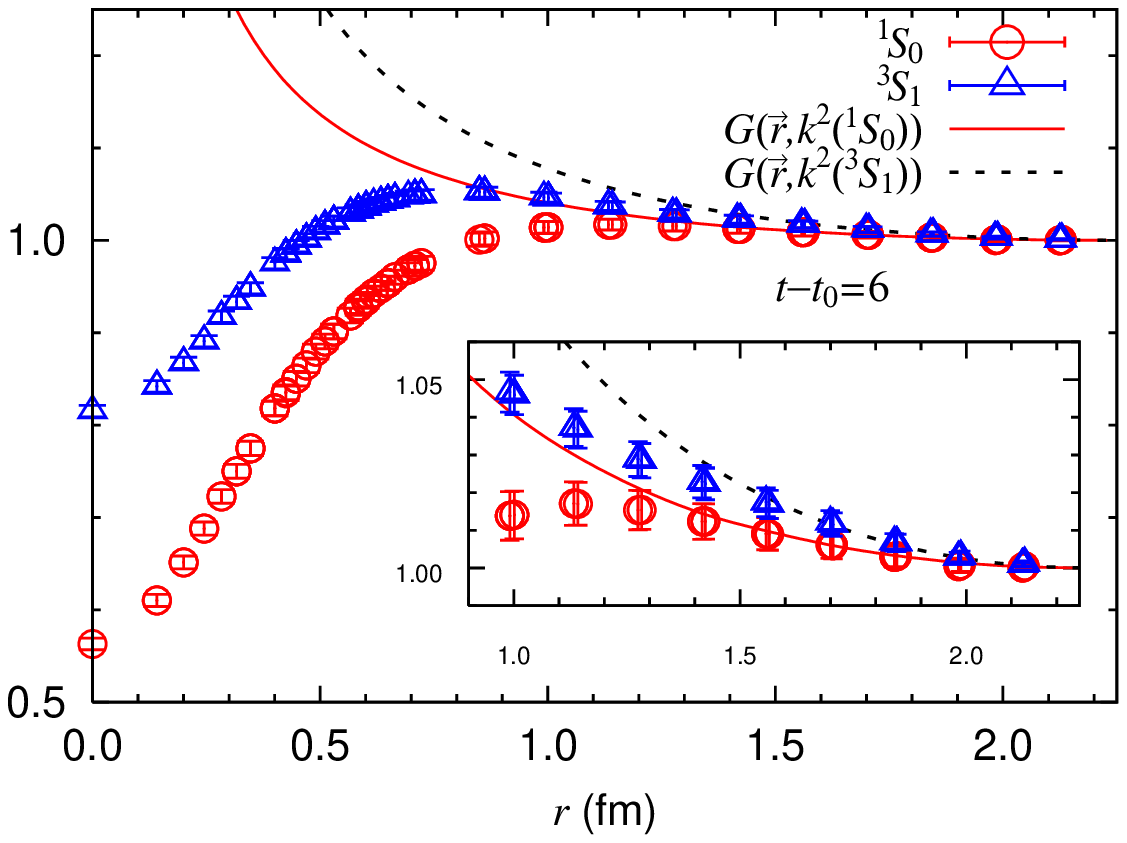}
 ~\hfil~
 \includegraphics[width=.490\textwidth]{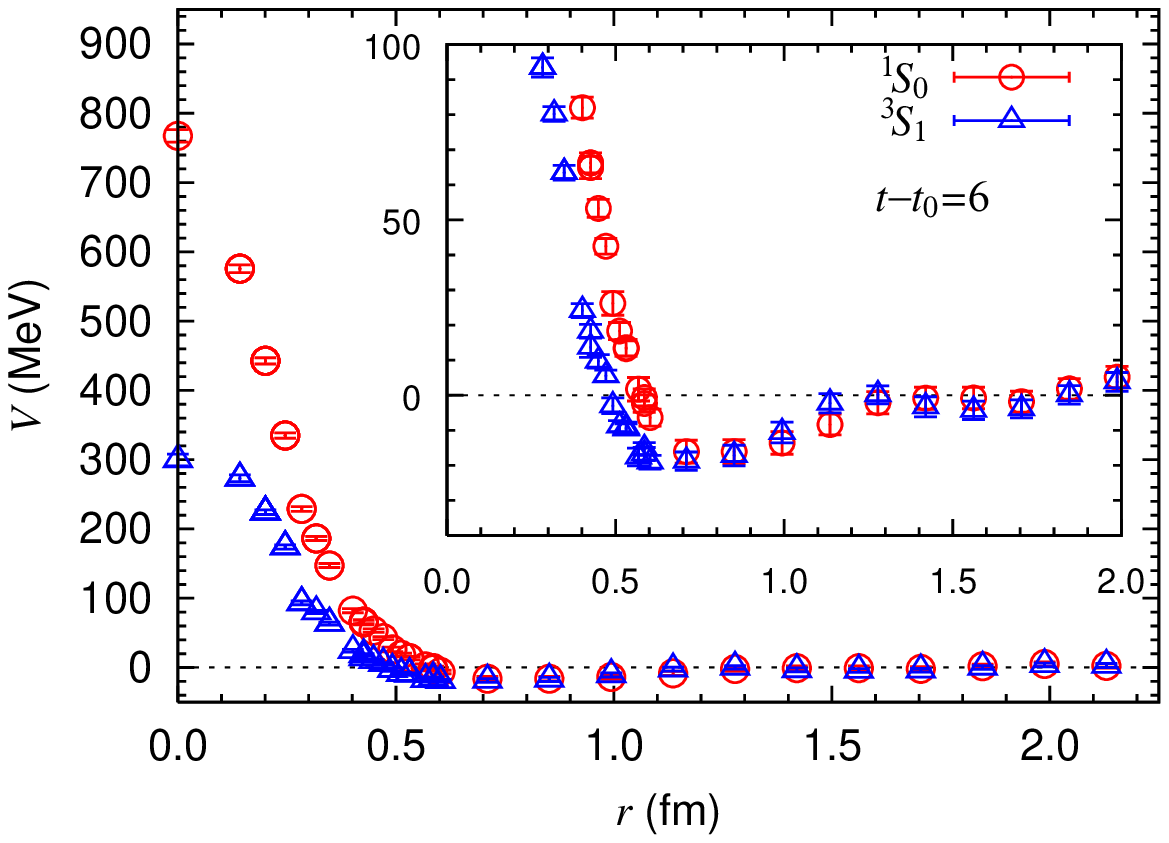}
 \caption{
 (Left)
 The radial wave function of $p\Xi^0$,
 in $^1S_0$ (circle) and $^3S_1$ (triangle) channels,
 obtained at $t-t_0=6$.
 The Green's functions $G(\vec{r},k^2)$
 with $\vec{r}=(r,0,0)$ 
 in $^1S_0$ (solid line) and
 $^3S_1$ (dotted line) are also shown.
 The inset shows its enlargement. 
 (Right)
 The effective central potential for $p\Xi^0$,
 in the $^1S_0$ (circle) and $^3S_1$ (triangle),
 obtained from the wave function at time slice $t-t_0=6$. 
 The inset shows its enlargement. 
 }
 \label{wavepot}
\end{figure}

The left panel of Fig.~\ref{wavepot} shows the wave functions 
obtained at time slice $t-t_0=6$ with 
$t_0= 5$. %
The open circle (triangle) corresponds to the case of the 
 $^1S_0$ ($^3S_1$) channel. 
They are normalized to be unity 
at the spatial boundary $\vec{r}=(32/2,0,0)$.  For  $r\lsim 0.7$ fm,
 the wave function is calculated for all possible values of $\vec{r}$,
while, for the outer region, 
it is calculated only on the $x, y, z$ axis and their nearest neighbors
 to reduce the numerical cost.

As seen in the Figure, the wave functions are suppressed at 
short distance and are 
slightly enhanced at 
medium distance in both $^1S_0$ and $^3S_1$ channels. 
There is a sizable difference 
of the suppression pattern at short distance 
between  $^1S_0$ and $^3S_1$; the repulsion is 
 stronger in  the $^1S_0$ channel.

\subsection{central potential}
\label{sec:central potential}

To derive the $N\Xi$ potential through Eq.~(\ref{EffCPot}),
we need to know the non-relativistic energy  $E=k^2/(2\mu)$.
 It has been shown in Ref.~\cite{CP-PACS} that $k^2$ and thus $E$ 
can be accurately determined by fitting the wave function 
 in the asymptotic region
 in terms of the lattice Green's function 
\begin{equation}
 G(\vec{r}, k^2) =
  {1\over L^3}
  \sum_{\vec{p}\in\Gamma}
  {1\over p^2 - k^2}
  {\rm e}^{i \vec{p}\cdot\vec{r}}, 
  \qquad
  \Gamma = \left\{
	    \vec{p};  \ \vec{p} = \vec{n}{2\pi\over L},
	    \vec{n}\in{\bfi{Z}^3}
	   \right\} ,
  \label{eq:Green}
\end{equation}
which is the solution of 
$(\bigtriangleup + k^2)G(\vec{r}, k^2)=-\delta_L(\vec{r})$
 with $\delta_L(\vec{r})$ being the periodic delta function 
 \cite{Luscher,CP-PACS}. 
Results of the fit in the range 
$(12 \le x \le 16, \ 0 \le y \le 1,\ z=0)$ are shown in the left panel of Fig.~\ref{wavepot}. 
The fitting range is determined so that it is 
 outside the range of the interaction (see the right panel of
 Fig.~\ref{wavepot}).
  The non-relativistic energies thus obtained 
 in the $^1S_0$ ($^3S_1$) channel 
 becomes $E=-0.4(2)$ MeV ($E=-0.8(2)$ MeV).
 We have checked that the results of the fit in different ranges
 $11 \le x \le 16$ and $13 \le x \le 16$ introduce 
 systematic errors only less than half of the statistical errors for
 $E$.
 Note that $E$ can be negative for the scattering state in a 
 finite box if the interaction is attractive.

\begin{figure}[t]
 \centering \leavevmode
 \begin{minipage}[t]{0.49\textwidth}
  \includegraphics[width=.94\textwidth]{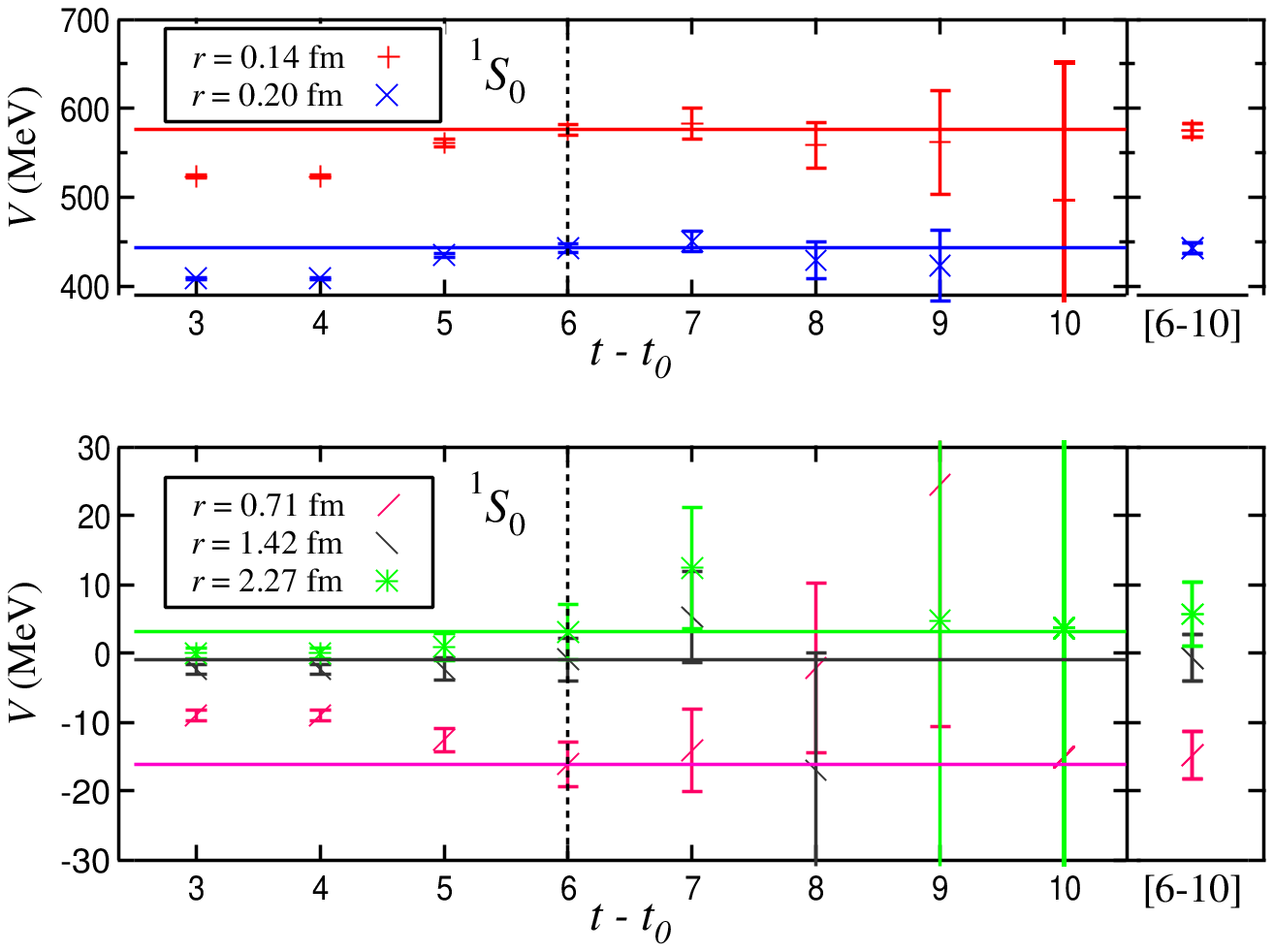}
  \footnotesize 
 \end{minipage}
 \hfill 
 \begin{minipage}[t]{0.49\textwidth}
  \includegraphics[width=.94\textwidth]{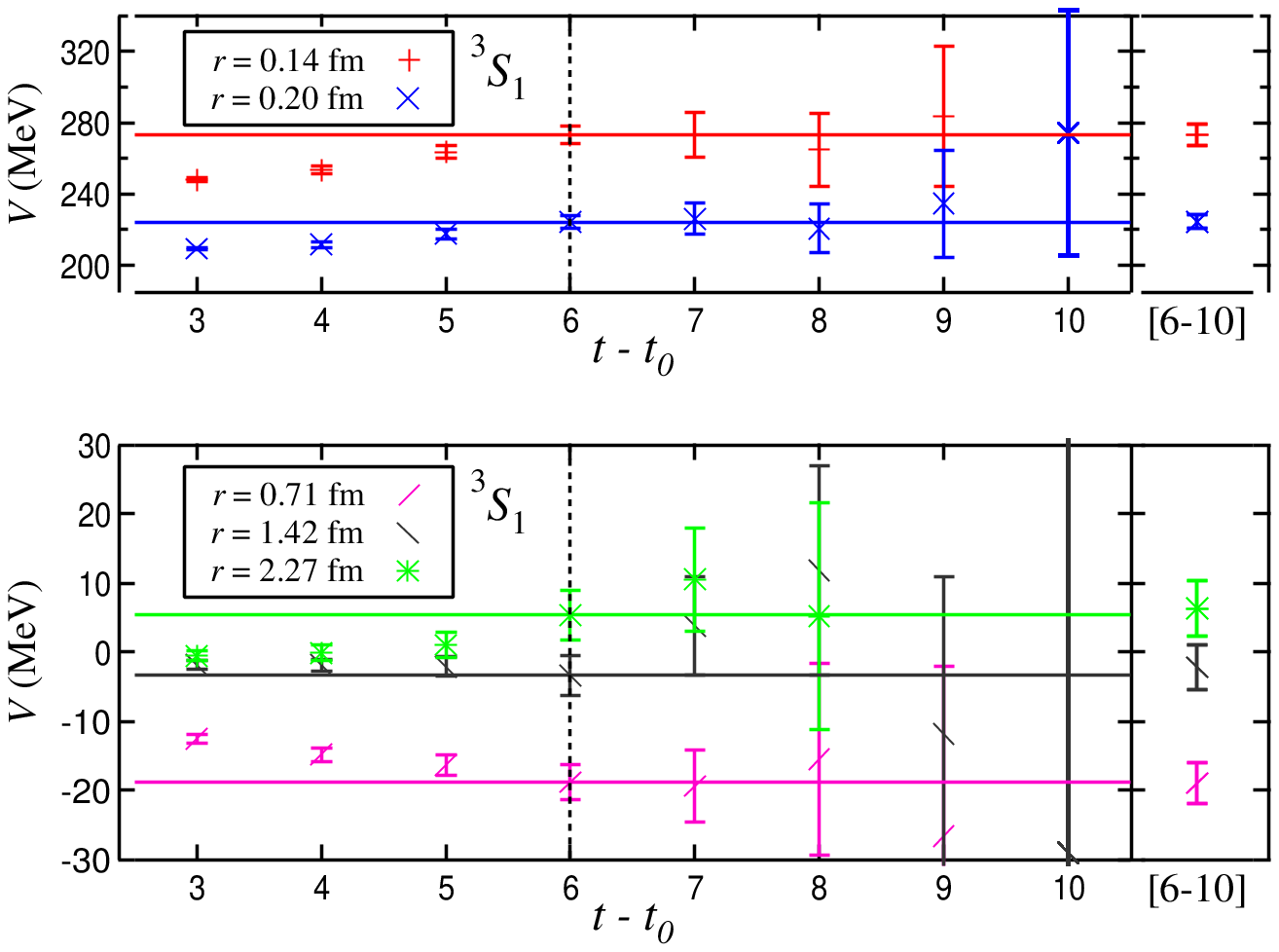}
 \end{minipage}
 \caption{The $t$ dependence of the potential 
 in the $^1S_0$ channel (left panel) and the $^3S_1$ channel (right panel) 
 for several values of $r$.
 The horizontal lines denote 
 the central values at $t-t_0=6$ which are adopted for the potential in Fig.1
  (right).  The results indicated by $[6-10]$ are obtained by
 a constant fit to the data for $t-t_0=6-10$
 with a single-elimination of the jackknife method.
 }
 \label{pots}
\end{figure}

The effective central potentials for $p\Xi^0$ system 
obtained from Eq.~(\ref{EffCPot}) with the wave function $\phi$ and the
 energy $E$  at $t-t_0=6$ are shown
 in the right panel of Fig.~\ref{wavepot} for
  the $^1S_0$ and $^3S_1$ channels at $m_{\pi}\simeq 368$ MeV. 
In order to check whether the ground state saturation of the $p\Xi^0$ system
 is achieved in the present results, 
 we plot the $t$-dependence of the potential 
 at distances $r=0.14$, $0.20$, $0.71$, $1.42$ and $2.27$ fm
 in Fig.~\ref{pots}.
 The fact that
the potential is stable against $t$ for $t-t_0 \ge  6$ within errors
indicates that
the ground state saturation is indeed achieved at $t-t_0=6$.
 This is the reason why
 we adopted the values at $t-t_0=6$ in the right panel of Fig.\ref{wavepot}.
 Note that a constant fit to data for $t-t_0 = 6-10$ with a 
 single elimination of the jackknife method 
  does not introduce appreciable change of the final values and the errors  
 of the potential as shown in Fig.\ref{pots}.

In  the right panel of Fig.~\ref{wavepot}, the potential  in the $N \Xi $ system
shows a repulsive core at $r \lsim 0.5$ fm surrounded by an attractive well,
similar to the $NN$ system \cite{Ishii:2006ec,Ishii:2007xz}.
In contrast to the $NN$ case, however,
one finds that
the repulsive core of the $p\Xi^0$ potential in the $^1S_0$ channel
is substantially stronger than that in the $^3S_1$ channel.
Such a large spin dependence is also suggested by 
the quark cluster model \cite{Oka:2000wj}.
The relatively weak attraction in the medium to
long distance region that ($0.6$ fm $\lsim r \lsim 1.2$ fm) 
 is similar in both $^1S_0$ and $^3S_1$ channels. 
As the energies determined above and scattering lengths
given later indicate, 
the present potentials are weakly attractive on the whole
in both spin channels in spite of 
the repulsive core at short distance. Also, 
 the attraction $^3S_1$ channel seems a little stronger than that 
in the $^1S_0$ channel.

\subsection{quark mass dependence}

\begin{figure}[t]
 \centering \leavevmode
 \includegraphics[width=.45\textwidth]
 {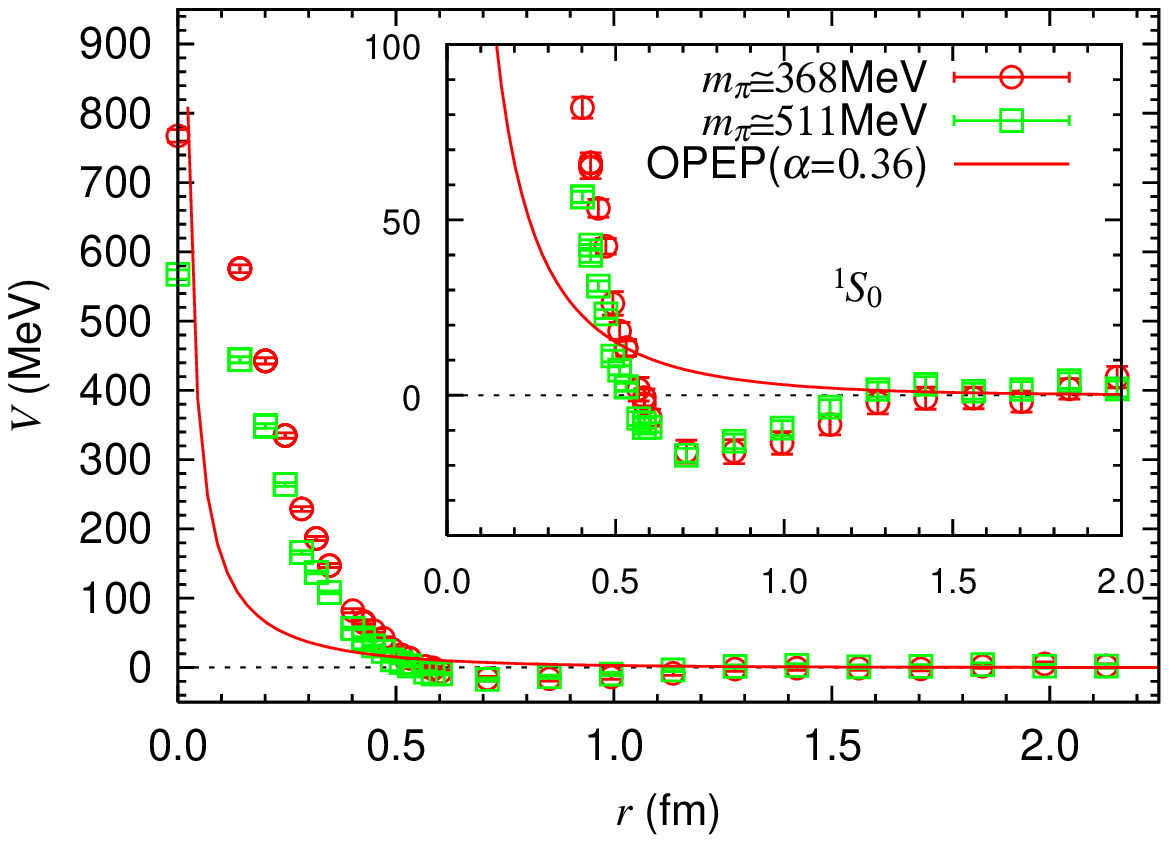}
 ~\hfil~
 \includegraphics[width=.45\textwidth]
 {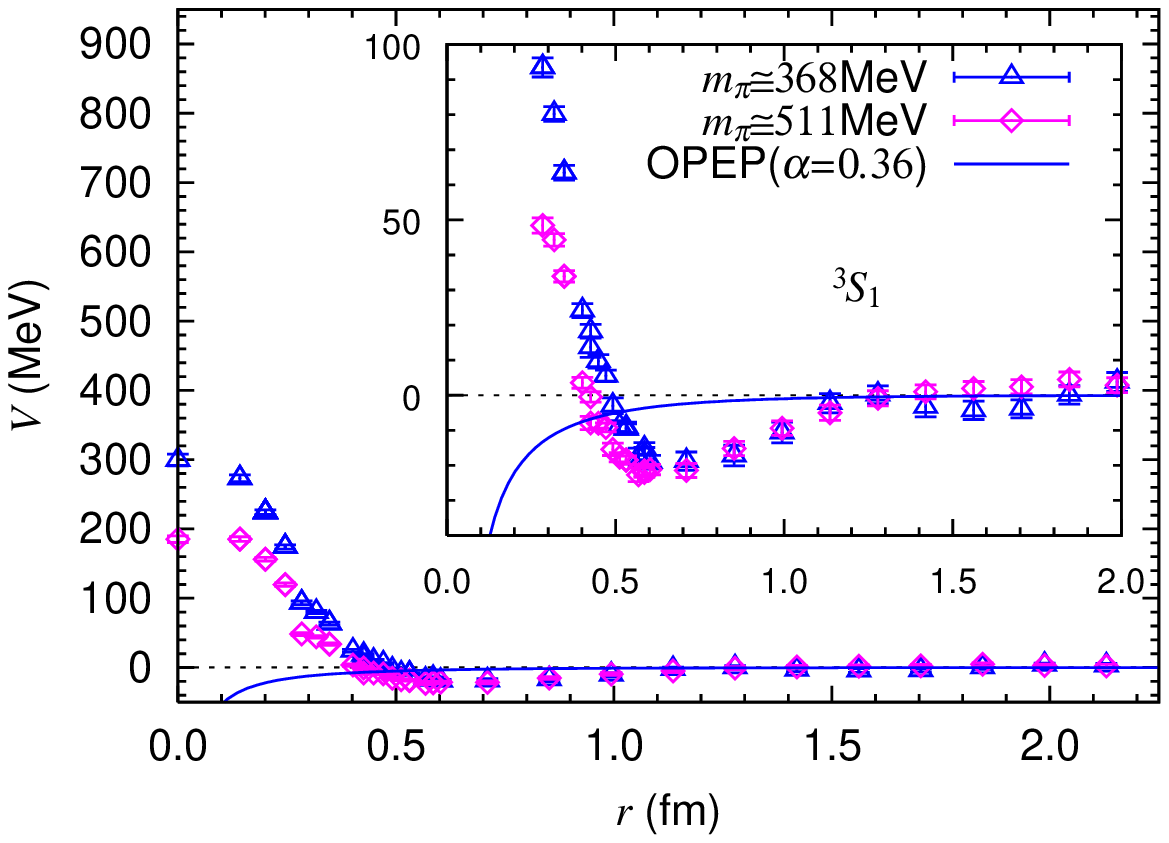}
 \caption{
 (Left)
 The effective central potential for $p\Xi^0$
 in the $^1S_0$ channel at $m_\pi\simeq 368$ MeV
 (circle) and $m_\pi\simeq 511$ MeV (box). 
 The central part of the OPEP ($F/(F+D)=0.36)$ in
 Eq.~(\ref{eq:opep}) is also given by solid line. 
 (Right)
 Same as the left figure, but  in the $^3S_1$ channel at 
 $m_\pi\simeq 368$ MeV (triangle) and 
 $m_\pi\simeq 511$ MeV (diamond). 
 }
 \label{fig:potqmass_w.opep}
\end{figure}
Figure~\ref{fig:potqmass_w.opep} compares the $p\Xi^0$ potential
at $m_\pi\simeq 368$ MeV with 
that at $m_\pi\simeq 511$ MeV
in the $^1S_0$ channel (left) and in the $^3S_1$ channel (right). 
At $m_\pi\simeq 511$ MeV,
the potentials are evaluated
at $t-t_0=7$. The non-relativistic energy in this case is
$E=-0.19(4)$ MeV ($-0.34(4)$ MeV) in the $^1S_0$ ($^3S_1$) channel, 
where the region that
$(10\le x\le 16,0\le y\le 1,z=0)$ is adopted to
 fit the wave function by the Green's function. 
 We have checked that the results of the fit in different ranges
 $9 \le x \le 16$ and $11 \le x \le 16$ introduce 
 systematic errors only less than half of the statistical errors for
 $E$.
 
The height of the repulsive core increases as the $ud$
 quark mass decreases, while the significant difference
 is not seen in the medium to long distances within the 
  error bars.

\subsection{one-pion exchange}

The exchange of a single  $\pi^0$ would conduct the interaction at
long distance in the $p\Xi^0$ system:
\begin{equation}
\label{eq:opep}
 V_C^{\pi} = - (1-2\alpha) \frac{g_{\pi NN}^2}{4\pi}
  {(\vec{\tau}_N\cdot\vec{\tau}_\Xi)
  (\vec{\sigma}_N\cdot\vec{\sigma}_\Xi)\over 3}
  \left(
   {m_\pi\over 2m_N}
  \right)^2
  {{\rm e}^{-m_\pi r}\over r},
\end{equation}
 The pseudo-vector $\pi NN$ coupling $f_{\pi NN}$ 
 and the $\pi \Xi \Xi$ coupling $f_{\pi \Xi\Xi}$ are
 related as $f_{\pi \Xi\Xi} = - f_{\pi NN} (1-2\alpha)$ 
 with the parameter $\alpha=F/(F+D)$ \cite{ESC04}.
 Also we define 
 $g_{\pi NN} \equiv f_{\pi NN} \frac{m_{\pi}}{2m_N}$.

The solid lines in Fig.\ref{fig:potqmass_w.opep}
 is the one pion exchange potential (OPEP) 
obtained from Eq.(\ref{eq:opep})
with $m_\pi\simeq 368$ MeV, $m_N\simeq 1167$ MeV 
(corresponding to $\kappa_{ud}=0.1678$)
 and the empirical values, $\alpha\simeq 0.36$
  \cite{Yamanishi:2007zza} and 
  $g_{\pi NN}^2/(4\pi)\simeq 14.0$ \cite{Machleidt:2001rw}.
 Unlike the $NN$ potential in the $S$-wave,
  the OPEP in the present case has opposite sign between
   the spin-singlet channel and the spin-triplet channel.
 Also, the absolute magnitude of OPEP is weak due to the 
  factor $1-2\alpha$.
  As is seen from Fig.\ref{fig:potqmass_w.opep},
  we do not find clear signature of OPEP at long distance
  ($r > 1.2$ fm) in our potential within statistical errors. 
 On the other hand, there is a clear departure from OPEP 
  at medium distance ($0.6 {\rm fm} < r < 1.2 {\rm fm}$)
  in both $^1S_0$ and $^3S_1$ channels.  These observations may indicate 
  a mechanism of 
   state-independent attraction such as the correlated two
   pion exchange.  
   
It should be mentioned  here that there is in principle 
a quenched artifact to the baryon-baryon potentials from the flavor singlet
hairpin diagram (the ghost exchange) \cite{Beane:2002nu}.
Its contribution to the central potential has a spin-dependent
 exponential tail  $V^{\eta}_{\rm C}(r) \propto
 \vec{\sigma}_{N}\cdot \vec{\sigma}_{\Xi} \exp(-m_{\pi}r)$, which 
 dominates over the Yukawa potential at large distances. 
 Its significance
can be estimated by comparing the sign and the magnitude
of $e^{m_{\pi}r} V_{\rm C}(r) $ in the spin-singlet and 
spin-triplet channels.
Our present data at lightest quark mass $m_{\pi}=368$ MeV
shows no evidence of the ghost contribution at large
distances within errors. This may  indicate the 
 weak coupling of the $\eta$ to $N$ and $\Xi$.

\subsection{scattering length}

\begin{figure}[t]
 \centering \leavevmode
 \includegraphics[width=.45\textwidth]{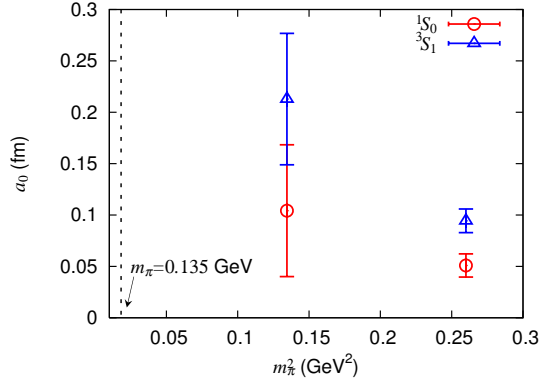}
 \caption{The scattering length for $p\Xi^0$
 in the $^1S_0$ (circle) and $^3S_1$ (triangle),
 obtained by the L\"{u}scher's formula from the asymptotic behavior of
 the wave function.
 The vertical dashed line is the physical point at $m_{\pi}=0.135$ GeV. 
 }
 \label{sctlng}
\end{figure}

Using the energy $E=k^2/(2\mu)$ obtained from the asymptotic behavior of 
the wave function in Sec.\ref{sec:central potential},
the $N\Xi$ scattering lengths can be deduced from
the L\"{u}scher's formula \cite{Luscher,CP-PACS},
\begin{equation}
 k \cot \delta_0(k) =
  \frac{2}{\sqrt{\pi} L} Z_{00}(1;q^2) =
  1/a_0 + O(k^2), 
\end{equation}
 where $Z_{00}(1;q^2)$ with $q=\frac{kL}{2\pi}$
 is obtained by the analytic continuation of the 
  generalized zeta-function $Z_{00}(s;q^2) =
   \frac{1}{ \sqrt{4\pi} } \sum_{{\vec n} \in {\bf Z}^3}
    (n^2-q^2)^{-s}$ defined for ${\rm Re}\ s > 3/2$. 
 The sign of the $S$-wave scattering length 
$a_0$ is defined to be positive for weak attraction.
 The results are plotted in Fig.\ref{sctlng} for two
  different values of the pion masses,
  $m_{\pi}\simeq 368$ MeV and $m_{\pi}\simeq 511 $MeV.
 Both $a_{0t}$ (the scattering length in the $^3S_1$ channel)
 and $a_{0s}$ (the scattering length in the $^1S_0$ channel)
slightly increase as the quark mass decreases:
 $a_{0s} =$ 0.05(1) $\rightarrow$  0.10(6) fm and $a_{0t} =$  0.09(1) $\rightarrow$ 0.21(6) fm. 
 The positive sign indicates that the $p\Xi^0$ interaction is 
 attractive on the whole in both $^1S_0$ and $^3S_1$ channels. 
 Due to the possible non-linear dependence of the scattering length as a function of  the quark mass
 as is well-known in the case of the $NN$ system
 \cite{Kuramashi:1995sc} (see also \cite{NPLQCD,Epelbaum:2005pn}
 and references therein),
  we are not able to obtain a reliable estimate of the scattering length 
 of the $p\Xi^0$ system at the physical $ud$ quark mass.
 In addition,
a possible contamination from the ghost exchange may arise 
at small quark masses in the quenched QCD simulations \cite{Beane:2002nu}.
Therefore, it is necessary in the future to carry out the full QCD calculation
toward lighter quark masses to predict the precise value of $a_0$. 
   
It would be interesting here to summarize 
 diverse results on the $N\Xi$ scattering length in  other approaches. 
The $p\Xi^0$ interaction in 
chiral effective field theory \cite{Haidenbauer:2007ra}
predicts weak repulsive scattering lengths of 
$a_{0s} \sim -0.2$ fm and  $a_{0t} \sim -0.02$ fm. 
The phenomenological boson exchange model (e.g., SC97f)
\cite{ESC04} gives
$a_{0s} =-0.4$ fm and $a_{0t} = 0.030$ fm. 
The quark cluster model (fss2) \cite{Fujiwara:2006yh} gives
$a_{0s}= -0.3$ fm and $a_{0t}=0.2$ fm, while 
QCD sum rules \cite{Kondo} gives
$a_{0s}=3.4\pm 1.4$ fm and $a_{0t} = 6.0\pm 1.4$ fm.

\section{Summary}
\label{sec:summary}

We study the $p\Xi^0$ interaction in the $^1S_0$ and $^3S_1$ channels 
through the equal-time Bethe-Salpeter amplitude measured by the 
quenched lattice QCD simulations
on a ($4.5$ fm)$^4$ lattice with the quark masses corresponding to 
$m_\pi/m_\rho \simeq 0.45$ and $0.59$. 
We adopt specific choice of the baryon interpolating operators  
as given in Eqs.~(\ref{proton}) and (\ref{Xi}) and extract the potential
associated with these operators.
 The effective central potential deduced from the Bethe-Salpeter wave function 
has  repulsive core 
at short distance surrounded by attractive well at 
the medium and long distances. 
The scattering lengths 
determined by fitting the asymptotic wave function with
the solution of the Helmholtz equation
indicates that the $p\Xi^0$ interaction is attractive on the whole in 
both $^1S_0$ and $^3S_1$ channels. There is a slight tendency that
the $^3S_1$ interaction is more attractive than that in the $^1S_0$ channel.

 To reduce the uncertainties due to the lattice discretization
 at short distances, we need to carry out the simulations
  with smaller lattice spacing and/or with the improved lattice action.
Systematic studies in various other channels 
such as $\Lambda N$, $\Sigma N$ and  $\Lambda\Lambda$
are not only interesting by themselves but also  important 
for studying the structure of hypernuclei and  
 the interior of neutron stars. 
Also, (2+1)-flavor QCD simulations with realistic  quark masses will be the
 ultimate goal to make qualitative predictions of the $NN$ and $YN$ 
  interactions: Studies along this line with the PACS-CS gauge
  configurations\cite{PACSCS}
   are now under way.

\ack
This work is supported by the Large Scale Simulation Program No.07-07 (FY2007)
 of High Energy Accelerator Research Organization (KEK).
We are grateful for authors and maintainers of {\tt CPS++}\cite{cps},
of which a modified version is used for measurement done in this work.
H.~N. is supported by the Special Postdoctoral Researchers Program 
at RIKEN. 
This research was partly supported by Grants-in-Aid for Young Scientists 
(B) (No. 17740174) from the Japan Society for Promotion of Science
(JSPS), and by the Ministry of Education, 
Science, Sports and Culture, Grant-in-Aid 
(Nos. 13135204, 15540254, 18540253, 19540261, 
20340047).

\end{document}